\documentclass{article}
\usepackage{geometry}
\usepackage{authblk}
\usepackage{graphicx}
\usepackage{epsfig}
\usepackage{amsfonts}
\usepackage{amssymb}
\usepackage{amsmath}
\usepackage{epsf}
\usepackage{color}

\newcommand{\insertplot}[5]{\begin{figure}
 \hfill\hbox to 0.05in{\vbox to #5in{\vfill
 \inputplot{#1}{#4}{#5}}\hfill}
 \hfill\vspace{-.1in}
 \caption{#2}\label{#3}
 \end{figure}}
 \newcommand{\inputplot}[3]{
 \special{ps: plotfile #1}
}
\newcounter{fig}

\newcommand{\ee}{\end{equation}}
\newcommand{\eea}{\end{eqnarray}}
\newcommand{\be}{\begin{equation}}
\newcommand{\bea}{\begin{eqnarray}}

\begin{document}

\title{Note on super-critical charged boson stars}

\author[1]{Yves Brihaye \footnote{yves.brihaye@umons.ac.be}}
\author[2]{Betti Hartmann \footnote{b.hartmann@ucl.ac.uk}}

\affil[1]{Physique de l’Univers-Champs et Gravitation, Universit\'e de Mons, 7000 Mons, Belgium   }
\affil[2]{Department of Mathematics, University College London, Gower Street, London, WC1E 6BT, UK}

\date{\today}

\maketitle

\begin{abstract} 
We study the transition of charged boson stars from sub- to super-criticality.
This transition is defined as that choice of coupling constants  for which the Coulomb repulsion of two individual bosons (that make up the star) exactly cancels their gravitational attraction.
It was recently shown that without self-interaction  super-critical boson stars are unstable to decay into their individual constituents. 
Here we show that this is no longer true for the self-interacting case
and that boson stars can possess spatial oscillations in the scalar field. 
We also discuss the corresponding black hole solutions that carry charged scalar hair.
\end{abstract}
 
 \maketitle
\section{Introduction}
Boson stars \cite{kaup,misch,flp,jetzer,new1} are essentially macroscopic Bose-Einstein condensates that
self-gravitate. The complex scalar field model possesses a U(1) symmetry that leads to a globally conserved Noether charge which can be interpreted as the number of scalar bosons that make up the star. Self-interaction of the scalar field is not necessary for these compact objects to exist, but allows to have a flat space-time limit, the so-called $Q$-ball \cite{qballs1}. This non-topological soliton is interesting, in particular, in supersymmetric extensions of the Standard Model of Particle Physics as scalar fields appear naturally \cite{qballs2}. In \cite{Copeland:2009as} 
$Q$-balls were studied in a model arising from gauge-mediated supersymmetry breaking and an exponential self-interaction potential for the scalar field was used. 

Boson stars have a harmonic time-dependence of the phase of the complex valued scalar field and if the phase depends only on time, the corresponding energy-momentum tensor and with it the space-time are static. However, boson stars can rotate and then possess an angular momentum that is  an integer multiple of the Noether charge \cite{Kleihaus:2005me, Kleihaus:2007vk}. In this latter case, the phase of the scalar field depends also on the azimuthal angle. 

The U(1) symmetry can be gauged and the resulting boson stars carry electric charge that is equal to the product of Noether charge and gauge coupling constant, i.e. one can think of a charged boson star as made up of scalar bosons that each carry a charge equal to the gauge coupling.

Charged boson stars have been addressed first in a U(1) gauged scalar field
model with a fourth order self-interaction potential \cite{jetzer_vdb}. It was argued that solutions exist only up to that choice of
coupling constants for which the electric repulsion of two individual bosons
exactly cancels their gravitational attraction. Since then, charged boson stars have been studied for different self-interaction potentials: (a) for a V-shaped potential \cite{Arodz:2008nm, Kleihaus:2009kr}, (b) a 6th order potential \cite{Brihaye:2021mqk}, and (c) an exponential potential \cite{Brihaye:2014gua}. 
In  \cite{Pugliese:2013gsa} so-called mini boson stars, i.e. boson stars that possess no self-interaction of the scalar field have been studied and it was argued that super-critical boson stars,
i.e. boson stars for which the Coulomb repulsion exceeds the gravitational
attraction should exist. This was recently shown to be correct \cite{Lopez:2023phk}, but it was demonstrated that these super-critical boson stars are unstable to decay into their individual bosonic constituents.

In this note, we extend the study of \cite{Lopez:2023phk} to include the self-interacting case and 
show that super-critical boson stars can be stable with respect to the decay mentioned above. We also demonstrate how the domain of existence changes. Moreover, we investigate the corresponding black hole solutions that carry a cloud of charged scalar fields.
These solutions have been discussed for the first time in \cite{herdeiro_radu, Hong:2020miv} and consequently studied in more detail in \cite{brihaye_hartmann,brihaye_console_hartmann}. Here, we discuss a few additional
features in the context of super- and sub-criticality of these solutions in order to shed further light on the globally regular case. 

The model and field equations are given in Section II, while our numerical
results for boson stars are summarized in section III. In Section IV we briefly discuss the corresponding black hole solutions and we conclude in Section V.

\section{The model}
The action of the $4$-dimensional gravity-gauge-scalar field model reads:
\be
\label{action}
S=\int \sqrt{-g} {\rm d}^4 x \ {\cal L}
\ee
with Lagrangian density given by
\be 
{\cal{L}} = \frac{R}{16 \pi G} - D_{\mu} \Psi^{\dagger}  D_{\mu} \Psi - U(|\Psi|) - \frac{1}{4} F^{\mu \nu} F_{\mu \nu} 
\ee
where $R$ is the Ricci scalar, $G$ Newton's constant, $\Psi$ is a complex-valued scalar field with potential $U(|\Psi|)$.
$D_{\mu} = \partial_{\mu} + i e A_{\mu}$ is the covariant derivative operator and
$F_{\mu \nu}=\partial_{\mu} A_{\nu} - \partial_{\nu} A_{\mu}$ the field strength tensor of a U(1) gauge field. 
Very frequently in the construction of boson stars, a 6-th order potential is chosen \cite{Kleihaus:2005me,Kleihaus:2007vk}~:
\begin{equation}
U_6(\psi) = \mu^2 \psi^2 - \lambda \psi^4 + \nu \psi^6
\end{equation}
 where $\mu$ is the mass of the scalar field and $\lambda$ and $\nu$ are positive constants that have to be chosen appropriately. An exponential potential, which has been first discuss in
the context of gauged supersymmetry breaking models \cite{Copeland:2009as} has also been used previously \cite{Brihaye:2014gua}~:
	\bea
	\label{susy_pot}
	    U_{\rm SUSY}(\psi) &=& \mu^2 \eta^2 \left(1 - \exp(-\psi^2/\eta^2)\right) \ \  ,
	\eea
 where $\mu$ is the mass of the scalar field and $\eta$ an energy scale. 
 
We would like to  discuss stationary, spherically symmetric solutions to the field equations resulting from
the variation of the action associated to (\ref{action}). 
To simplify  these equations we use a spherically symmetric Ansatz for the metric and matter fields~:
\begin{equation}
\label{ansatz}
{\rm d}s^2 = -(\sigma(r))^2 N(r) {\rm d}t^2 + \frac{1}{N(r)} {\rm d}r^2 + r^2\left({\rm d}\theta^2 + \sin^2 \theta {\rm d}\varphi^2 \right) \ \ , \ \   A_{\mu} {\rm d} x^{\mu} = V(r) {\rm d} t  \ \ , \ \ 
\Psi=\psi(r) \exp(-i\omega t) \ .
\end{equation} 
In the following, we will define a mass function as follows
\begin{equation}
N(r)=1 - 2\frac{m(r)}{r} \ .
\end{equation}
Substituting  the Ansatz (\ref{ansatz}) into the equations of motion, we find with $U_{(X)}=U_6$ or $U_{(X)}=U_{\rm SUSY}$, respectively, that 
\bea
    m' &=& 4\pi G r^2 \biggl[ \frac{V'^2}{2 \sigma^2} + N \psi'^2 + U_{(X)}(\psi) + \frac{((\omega-e V) \psi)^2}{N \sigma^2} \biggr] \ ,  \\ 
		\sigma' &=&  8\pi G  r \sigma \biggl[ \psi'^2 + \frac{((\omega-e V) \psi)^2}{N^2 \sigma^2} \biggr] \ ,
\label{equations_gr}
\eea
for the metric functions and 
 \be
	V'' 
+  \biggl[ \frac{2}{r} -\frac{\sigma'}{\sigma} \biggr] V' 		+ \frac{2 e(\omega - e V) \psi^2}{N} = 0
\label{equation_1}
		\ee
		\be
			\psi'' +  \left(\frac{2}{r} + \frac{N'}{N} +\frac{\sigma'}{\sigma}\right) \psi' + \frac{(\omega -e  V)^2 \psi}{N^2 \sigma^2} - \frac{1}{2N}
			\frac{dU_{(X)}}{d \psi} = 0
\label{equation_2}
\ee
for the matter field functions. Note that due to the U(1) gauge symmetry, the field equations depend only on the gauge invariant combination $\omega - e V(r)$. 

Using the following rescalings
\be
       r\rightarrow \frac{r}{\mu} \ \ \ , \ \ \ m(r) \rightarrow \frac{m(r)}{\mu} \ \ \ , \ \ \  \psi\rightarrow \eta \psi \ \ \ , \ \ \ V\rightarrow \eta V \ \ \ , \ \ \  \omega \rightarrow \mu \omega
\ee
leads to the observation that the equations depend only on the following two dimensionless coupling constants
\be
\alpha=8\pi G \eta^2 \ \ \ , \ \ \  q=\frac{\eta}{\mu} e  \ .
\ee 
Note that in the case of a pure mass potential, we can apply another
scaling and set $\alpha\equiv 1$ without loss of generality.

In the following we will study boson stars as well as black holes solutions of the set of coupled non-linear ordinary differential equations (\ref{equations_gr}), (\ref{equation_1}), (\ref{equation_2}). We require these solutions to be  asymptotically flat and hence require~:
\be
      \sigma(r \to \infty) \rightarrow 1    \ \ , \ \ \phi(r \to \infty) \rightarrow 0 \ \ .
\ee
From the asymptotic behaviour of the solutions we can read off the  physical quantities. These are the mass $M$ and electric charge $Q$~:
\begin{equation}
\label{eq:infty}
N(r \gg 1)=1-\frac{2M}{r} + \frac{\alpha Q^2}{2 r^2} + ..... \ \ , \ \ 
V(x)=v_{\infty}  - \frac{Q}{r} + .... \ .
\end{equation}
The solutions also possess a globally conserved Noether charge due to the U(1) symmetry. This reads
\begin{equation}
\label{eq:noether}
Q_N= \frac{1}{4 \pi}\int {\rm d}^3 x \ \sqrt{-g} j^0 \ \ = \ \int_{r_0}^{\infty} dr r^2 \sigma j^0 \ \ , \ \  
     \ \ j^0 =   \frac{2 (\omega-e V)}{N \sigma^2}\psi^2 \ .
\end{equation}
with $r_0=0$ for boson stars and $r_0=r_h$ for black holes. 
$Q_N$ can be interpreted as the number of scalar bosons making up the boson star or the cloud surrounding the black hole. 
For boson stars, i.e. globally regular solutions to the equations, we get $Q=qQ_N$, i.e. the total electric charge can be interpreted as made of $Q_N$ scalar bosons which each carry a charge $q$. For black holes, the total electric charge $Q$ is the charge contained in the cloud made of $Q_N$ scalar bosons each with charge $q$ plus the horizon electric charge given by $Q_H=V'(r_h)r_h^2/\sigma(r_h)$ \cite{herdeiro_radu}.

The asymptotic behaviour of the scalar field also determines the domain of existence of solutions. Indeed, from (\ref{equation_2}) we find that 
\begin{equation}
\label{eq:fall_off_scalar}
\psi(r\rightarrow \infty)\sim \frac{\exp(-\mu_{\rm eff,\infty} r)}{r} + .... \ \ , \  \
\mu^2_{{\rm eff},\infty}  =  \mu^2 - (\omega - e v_{\infty})^2 \ ,
\end{equation}
where $\mu_{\rm eff}$ is the effective mass of the scalar field which results from the difference between the ``bare'' mass $\mu$ and the electric potential energy. Obviously, we need to require $\mu^2_{{\rm eff},\infty} > 0$ in order to have an exponential decay of the solution. This means that the quantity
\be
\label{eq:Omega}
\Omega \equiv \mu - \omega + e v(\infty) 
\ee
needs to be positive. If $\Omega$ were negative, there would be enough electromagnetic energy to create scalar particles of mass $\mu$. \\

In \cite{jetzer_vdb} the transition from {\it sub- to super-criticality} was defined as choice of coupling constants such that the Coulomb force of two individual charged bosons making up the boson star exactly cancels their gravitational interaction. This means (in dimensionful units)
\begin{equation}
\frac{e^2}{4\pi r^2} = \frac{G \mu^2}{r^2} \ \ \ \ \Rightarrow \ \ \ \
q_{\rm cr}^2 = \frac{\alpha}{2}  \ .
\end{equation}

\section{Boson stars}
Boson stars are globally regular solutions to the equations (\ref{equations_gr}), (\ref{equation_1}), (\ref{equation_2}). We hence need to impose regularity conditions at $r=0$ which read 
\be
         N(0) = 1 \ \ , \ \ \psi'(0) = 0 \ \ , \ \ V(0) = 0 \ \ , \ \ V'(0) = 0  \ .
\ee
Note that the choice $V(0) = 0$ results from the fixing of the residual global symmetry.

\subsection{Numerical results}
We have solved the coupled, non-linear differential equations using a collocation solver \cite{COLSYS}.

\subsubsection{Scalar field without self-interaction}
\begin{figure}[h]
\includegraphics[width=8cm]{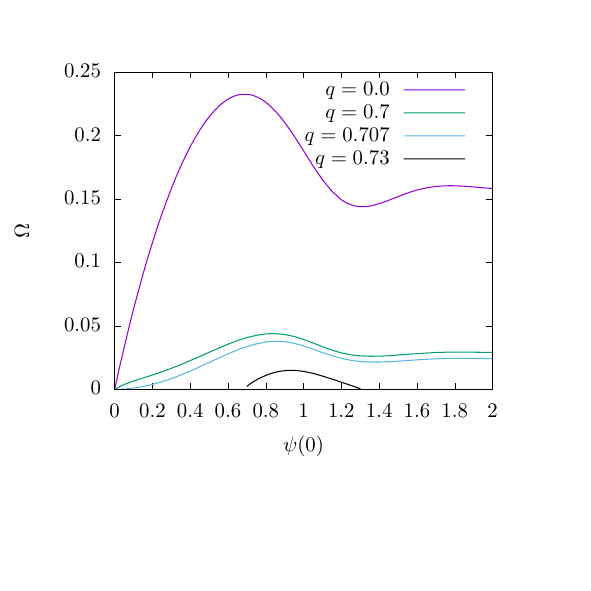}
\includegraphics[width=8cm]{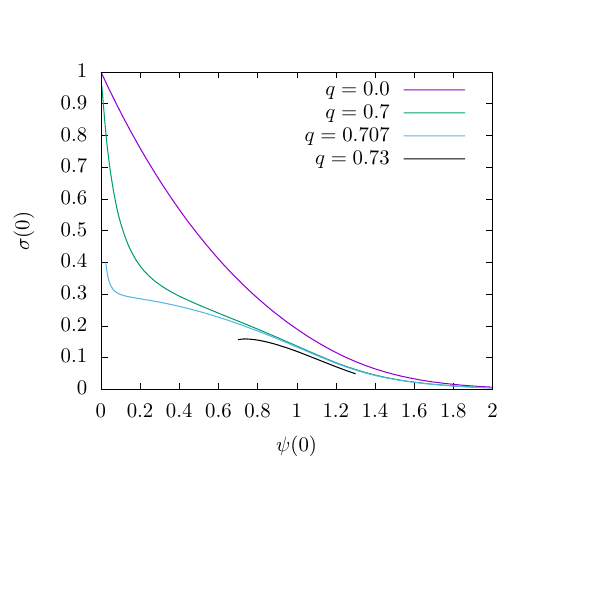}
\vspace{-2cm}
\caption{{\it Left}: The dependence of $\Omega$ on $\psi(0)$ for a massive scalar field without self-interaction and for several values of $q$ including $q\approx q_{\rm cr}=1/\sqrt{2}$ (blue).
{\it Right}: The dependence of $\sigma(0)$ on $\psi(0)$ for the same solutions. 
    \label{fig:mass_omega_sigma0}}
\end{figure}

In order to understand the influence of the self-interaction, we have first revisited the case of a mass term only, i.e. chosen $U_{(X)}=U_{\rm 6}$ with $\lambda=\nu=0$. In this case, we can set $\alpha\equiv 1$ (see discussion above). These solutions have been studied in  \cite{jetzer_vdb,Pugliese:2013gsa} and revisited recently in \cite{Lopez:2023phk}. As pointed out, solutions for $q\in [0:1/\sqrt{2}]$ (the sub-critical regime) are different to solutions for $q > 1/\sqrt{2}$ (the super-critical regime). In \cite{jetzer_vdb} it was shown that at $q=q_{\rm cr}=1/\sqrt{2}$ the Coulomb repulsion of the individual charged bosons making up the star exactly cancels their gravitational attraction.
However, as shown in \cite{Pugliese:2013gsa}, solutions exist for $q$ (slightly) above $q_{\rm cr}$ up to a (numerically given) maximal value $q_M$. This is due to the presence of the scalar field as well as the non-linear nature of the gravitational interaction and has recently been confirmed in \cite{Lopez:2023phk} and it was found that $q_M\approx 0.739$. Moreover, it was shown that the solutions for $q \in [q_{\rm cr}:q_M]$ are all unstable to decay into $Q_N$ individual bosons because $M > q Q_N$. In the following, we will demonstrate that the value of $q_M$ is connected to the vanishing of the parameter $\Omega$ (see (\ref{eq:Omega})) and can hence be determined quite precisely. In Fig. \ref{fig:mass_omega_sigma0} we show the dependence of $\Omega$ on $\psi(0)$ (left) as well as the dependence of $\sigma(0)$ on $\psi(0)$ (right). We find for
\begin{itemize}
    \item $q \in [0:q_{\rm cr}]$: solutions exist for $\psi(0)\in [0:\psi(0)_{\rm max}]$ where $\Omega$ vanishes at (and only at) $\psi(0)=0$ and $\Omega$ becomes constant for sufficiently large $\psi(0)$; our results indicate that $\psi(0)\rightarrow \infty$ and that in this limit $\sigma(0)\rightarrow 0$, i.e. an infinite central density of the star leads to a space-time singularity, as expected;
    \item $q \in [q_{\rm cr}:q_M]$: solutions exist only for $\psi(0)\in [\psi(0)_{\rm min}:\psi(0)_{\rm max}]$
    where $\Omega$ vanishes at both $\psi(0)_{\rm min} > 0$ and $\psi(0)_{\rm max} > 0$
    In fact, as $q$ is increased further we find that the interval $[\psi_{\rm min}(0):\psi_{\rm max}(0)]$ shrinks, i.e. the
maximal value of $\Omega$ up to where solutions exist decreases, too. At $q_M=0.73995$ we find that
$\psi_{\rm min}\approx\psi_{\rm max}\approx 0.97$, i.e. that the interval has shrunk to a point, and $\Omega=0$.
    At the same time, the metric functions $N(r)$ and $\sigma(r)$ remain perfectly finite. 
    \end{itemize}

\begin{figure}[h]
\includegraphics[width=8cm]{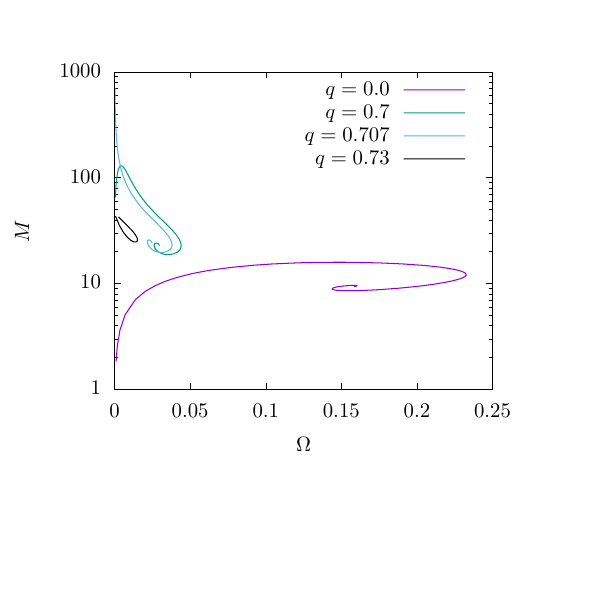}
\includegraphics[width=8cm]{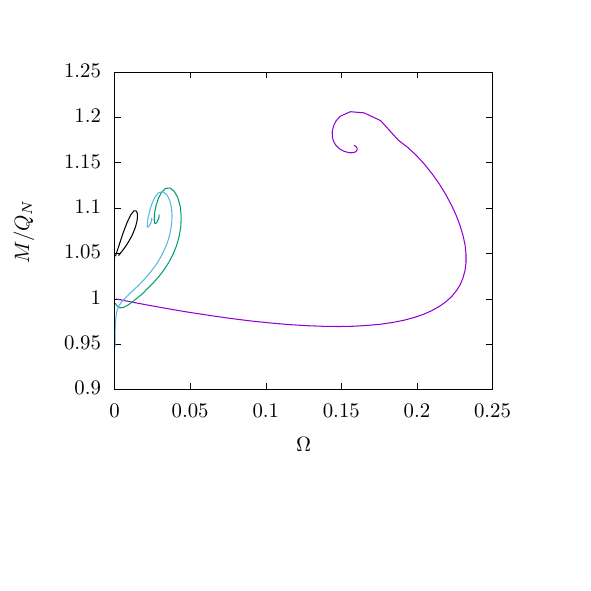}
\vspace{-2cm}
\caption{{\it Left}: The dependence of the mass $M$ on $\Omega$  for a massive scalar field without self-interaction and for several values of $q$ including $q\approx q_{\rm cr}=1/\sqrt{2}$ (blue).
{\it Right}: The dependence of the ratio $M/Q_N$ on $\Omega$ for the same solutions. 
    \label{fig:mass_ratio}}
\end{figure}

In Fig. \ref{fig:mass_ratio}, we give the mass $M$ of the solutions (left) as well as the ration $M/Q_N$ (right) in dependence on $\Omega$ for the same values of $q$. The mass $M$ shows the typical spiraling behaviour of boson stars
as long as $q \leq q_{\rm cr}$. When $q > q_{\rm cr}$ (here for $0.73$), we find a closed loop in the $\Omega-M-$ plane
that starts and finishes at $\Omega=0$. For $q=0.73$ we find that $\psi(0)_{\rm min}\approx 0.70$ and $\psi(0)_{\rm max}\approx 1.30$. The values of the ratio $M/Q_N$ demonstrate that super-critical solutions
are, indeed, unstable to decay into $Q_N$ individual bosons.
However, also sub-critical solutions can be unstable to this decay. For $q=0.0$, the solutions on the main branch are stable to this decay, while solutions on a part of the second and on further branches are unstable. Increasing $q$ from zero, also solutions on the main branch become unstable.

\subsubsection{Scalar field with exponential self-interaction}
We will now discuss the influence of the self-interaction on the above observations. For that we choose $U_{(X)}=U_{\rm SUSY}$. Note that these
solutions have been briefly discussed in \cite{Brihaye:2014gua}, but that details on super-critical solutions were not given. 

In contrast to the scalar field without self-interaction, solutions for $\alpha\rightarrow 0$ exist. These are non-topological solitons called {\it Q-balls}. In \cite{Brihaye:2014gua}, these solutions were investigated and it was shown that  they exist for a finite interval of $\psi(0) \in [\psi(0)_{\rm min}, \psi(0)_{\rm max}]$ where the values 
of $\psi(0)_{\rm min}$ and $\psi(0)_{\rm max}$ depend on $q$. At $\psi(0)_{\rm min}$ the electric field vanishes, while at $\psi(0)_{\rm max}$ it spreads over all space. This indicates that a minimal amount of bosonic particles
making up the star needs to be present for charged Q-balls to exist, while there can also not be too many as the electric repulsion takes over. In the limit  $q \to  q_M(\alpha)$  with $q_M(0) \sim 0.1262$ the interval in $\psi(0)$ shrinks to zero and no Q-ball solutions exist for $q > q_M(0)$. 

\begin{figure}[h]
\includegraphics[width=8cm]{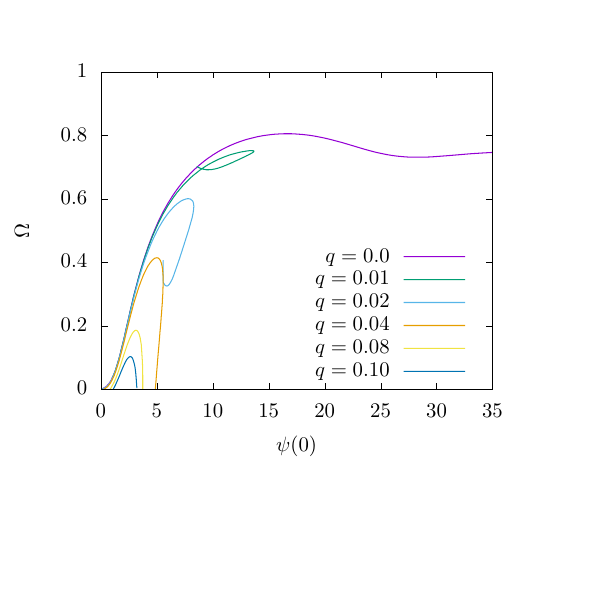}
\includegraphics[width=8cm]{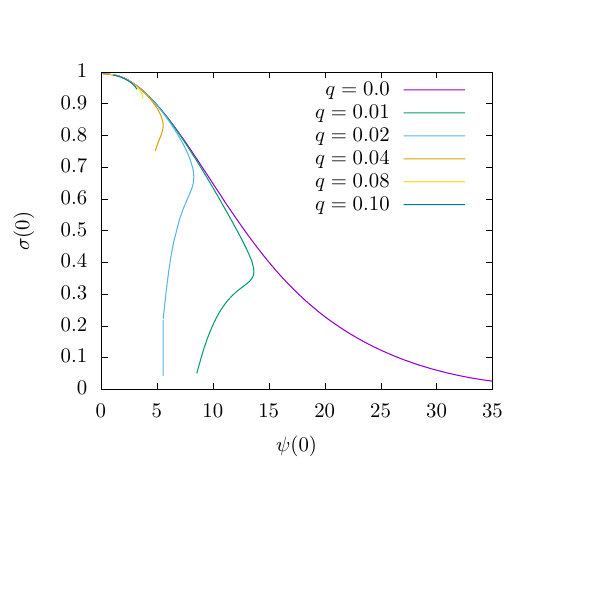}
\vspace{-2cm}
\caption{{\it Left}: The dependence of $\Omega$ on $\psi(0)$ for a massive scalar field with self-interaction,  for $\alpha=0.0012$ for several values of $q$.
{\it Right}: The dependence of $\sigma(0)$ on $\psi(0)$ for the same solutions. 
    \label{fig:mass_omega_sigma0_alp_0_0012}}
\end{figure}

\begin{figure}[h]
\includegraphics[width=8cm]{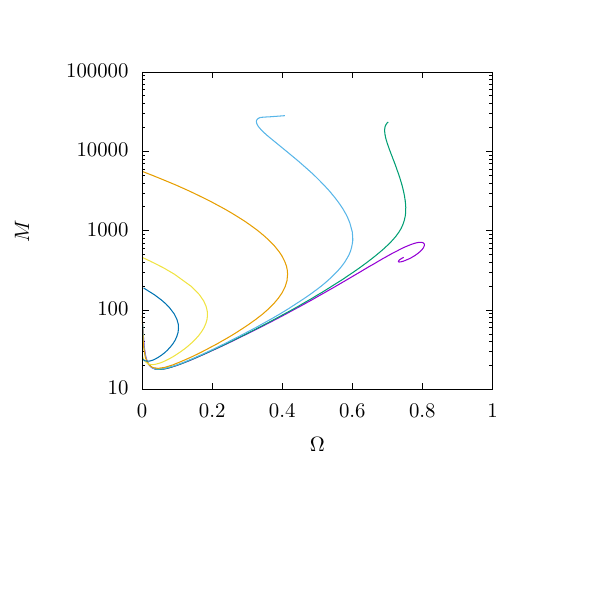}
\includegraphics[width=8cm]{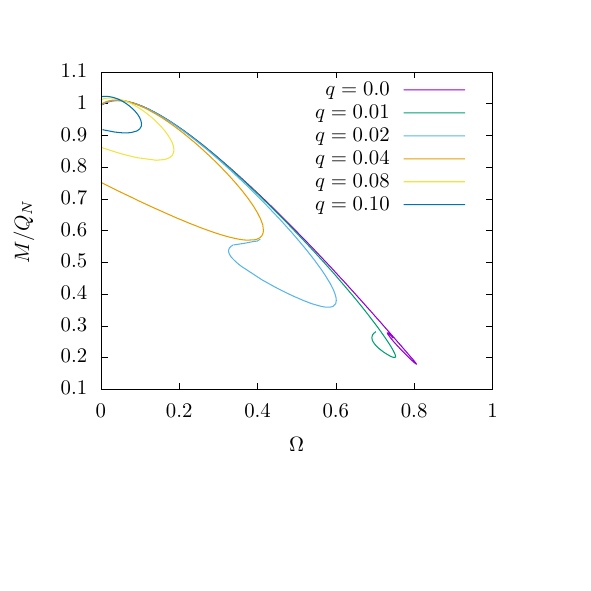}
\vspace{-2cm}
\caption{{\it Left}: The dependence of the mass $M$ on $\Omega$  for a massive scalar field with self-interaction, for $\alpha=0.0012$ and for several values of $q$.
{\it Right}: The dependence of the ratio $M/Q_N$ on $\Omega$ for the same solutions. 
    \label{fig:mass_ratio_alp_0_0012}}
\end{figure}

We have first studied the case $\alpha = 0.0012$ to understand how the coupling to gravity influences the value $q_M(\alpha)$. In this case, the value of $q$ which signals transition from sub- to super-criticality is $q_{\rm cr}(0.0012)\approx 0.0245$. Our results are shown in Fig. \ref{fig:mass_omega_sigma0_alp_0_0012}
and Fig. \ref{fig:mass_ratio_alp_0_0012}, respectively, and these should be compared to Fig. \ref{fig:mass_omega_sigma0} and Fig. \ref{fig:mass_ratio}.

We find that now three regimes of $q$ exist. These are
\begin{itemize}
\item $q=0$: this is the case of the uncharged boson star. The only energy scale is given in terms of $\alpha$, i.e. the ratio between the Planck mass and the energy scale of the self-interaction. $\psi(0)$ can become very large and in that limit $\sigma(0)\rightarrow 0$, while $\Omega$ tends to a finite value. The mass $M$ shows the typical spiraling behaviour in function of $\Omega$. Close to
$\Omega=0$ the ratio $M/Q_N$ can become larger than unity, indicating an instability to decay into individual bosons.

\item $ 0 < q\lesssim q_{\rm cr}(\alpha=0.0012)$:  these are the sub-critical, charged boson stars and these exist only on a finite interval of $\psi(0)\in [\psi_{\rm min}(0):\psi_{\rm max}(0)]$ with $\psi_{\rm min}=0$. At $\psi_{\rm max}(0)$
a second branch of solutions extends backwards in $\Omega$ and terminates
at a value of $\Omega_{\rm cr} > 0$ at finite $\psi(0)=\psi(0)_{\rm cr} > 0$ from where a third branch of solutions extends backwards for sufficiently large values of $q$. We find the third branch for $q=0.02$, but not for $q=0.01$. Solutions on this third branch of solutions have interesting new features that were first discussed in \cite{Brihaye:2021mqk}, albeit for a 6th order potential. In some intermediate region of the radial coordinate $r$, the scalar field develops
spatial oscillations. This can be understood when considering the scalar field
equation for small scalar field values which reads
\begin{equation}
\frac{\left(r^2 N \sigma \psi'\right)'}{r^2 N\sigma} = M_{\rm eff}^2 \psi \ \ \ , \ \ \ M_{\rm eff}^2(r) = N(r)^{-1} - \frac{q^2 V(r)^2}{N(r)^2 \sigma(r)^2} \ .
\end{equation}
For $q^2 V(r)^2$ sufficiently large and $N(r)$ sufficiently close to zero, the position dependent mass $M_{\rm eff}^2(r)$ can become negative leading to an oscillating behaviour of the scalar field function. This is shown in Fig. \ref{fig:oscillations}. As is obvious here, the solution splits into an ``inner'' and an ``outer'' part. 
The inside of the boson star has $\psi(r)\approx \psi(0) > 0$, i.e. the scalar field is more or less constant. Moreover,  $V(r)\equiv 0$. So, the interior is uncharged and contains a constant energy density. This can be interpreted as playing the role of a cosmological constant (compare \cite{Brihaye:2021mqk} for more details in the case of a $\psi^6$-potential).
This leads to a de Sitter horizon wanting to form, which is obvious from the behaviour of the metric function $N(r)$. However, it never quite reaches zero, and in the region just outside the core of the star, where $\psi(r)$ drops sharply, spatial oscillations of the scalar field appear.

\begin{figure}[h]
\includegraphics[width=8cm]{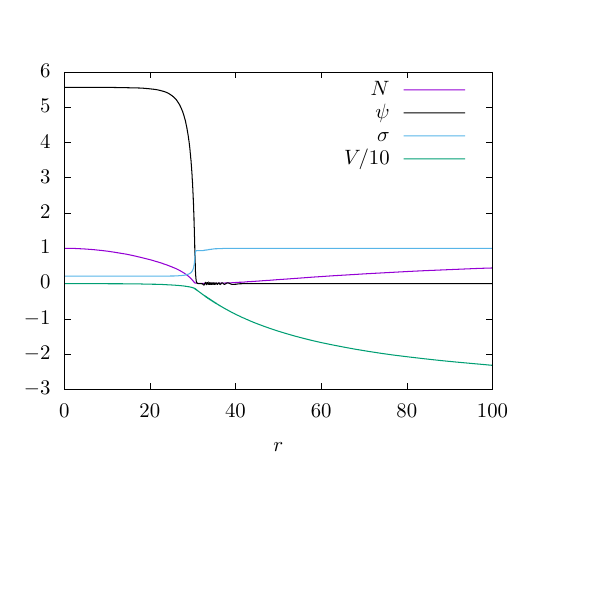}
\includegraphics[width=8cm]{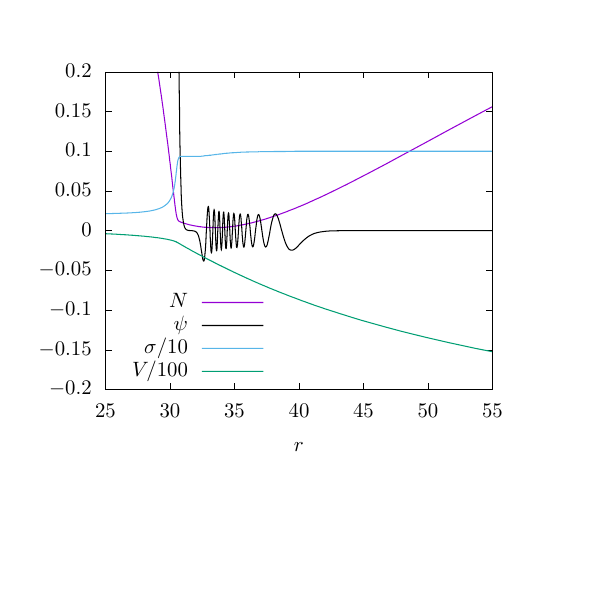}
\vspace{-2cm}
\caption{{\it Left}: We show the metric functions $N(r)$, $\sigma(r)$ as well as the scalar and gauge field functions
$\psi(r)$ and $V(r)$, respectively, for $\alpha = 0.0012$, $e=0.02$ and $\omega = 0.001$.
{\it Right}: Zoom into the region where oscillations of the scalar field function appear. \label{fig:oscillations}}
\end{figure}

Note that the solutions on the third branch have higher mass than those on the first and second, however, as Fig. \ref{fig:mass_ratio_alp_0_0012} demonstrates, these solutions are stable with respect to decay into individual bosons.

\item $q \gtrsim q_{\rm cr}(\alpha=0.0012)$: these are the super-critical boson stars. These solutions exist only on a finite interval of $\psi(0)\in [\psi_{\rm min}(0):\psi_{\rm max}(0)]$ with $\psi_{\rm min}(0)> 0$ and
$\Omega=0$ at both $\psi_{\rm min}(0)$ and $\psi_{\rm max}(0)$. We don't find
any spatially oscillating solutions in this case.
    
\end{itemize}

\begin{figure}[h]
\includegraphics[width=8cm]{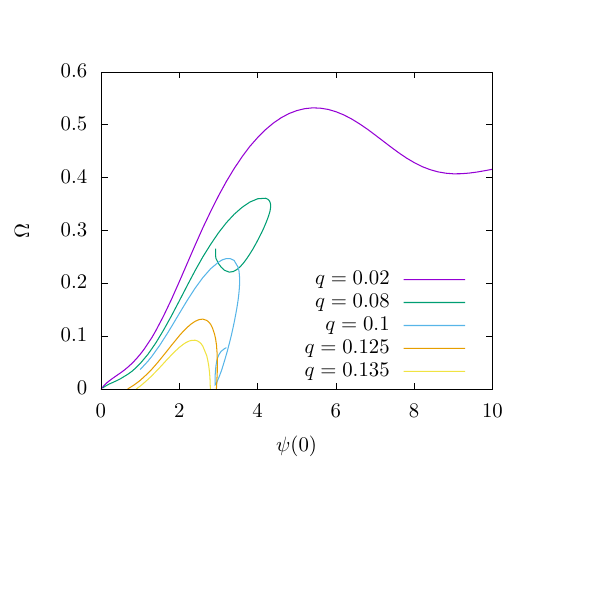}
\includegraphics[width=8cm]{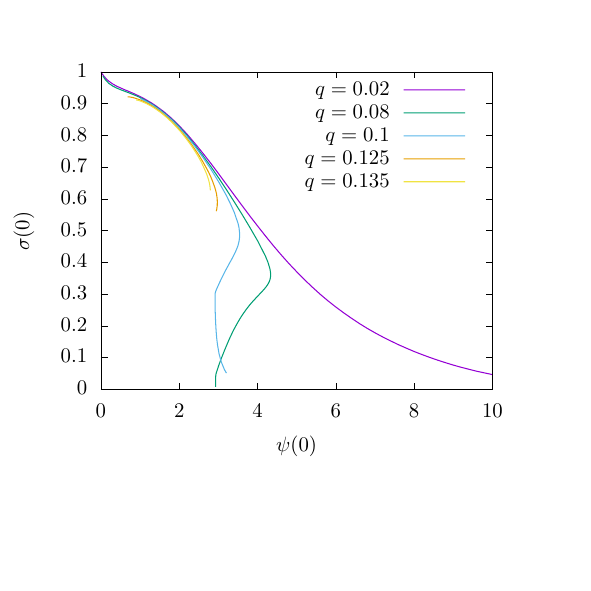}
\vspace{-2cm}
\caption{{\it Left}: The dependence of $\Omega$ on $\psi(0)$ for a massive scalar field with self-interaction,  for $\alpha=0.012$ for several values of $q$.
{\it Right}: The dependence of $\sigma(0)$ on $\psi(0)$ for the same solutions. 
    \label{fig:mass_omega_sigma0_alp0_012}}
\end{figure}  

\begin{figure}[h]
\includegraphics[width=8cm]{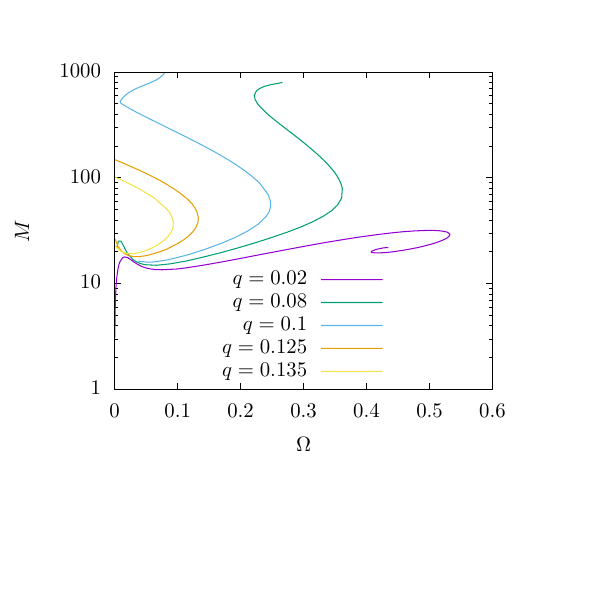}
\includegraphics[width=8cm]{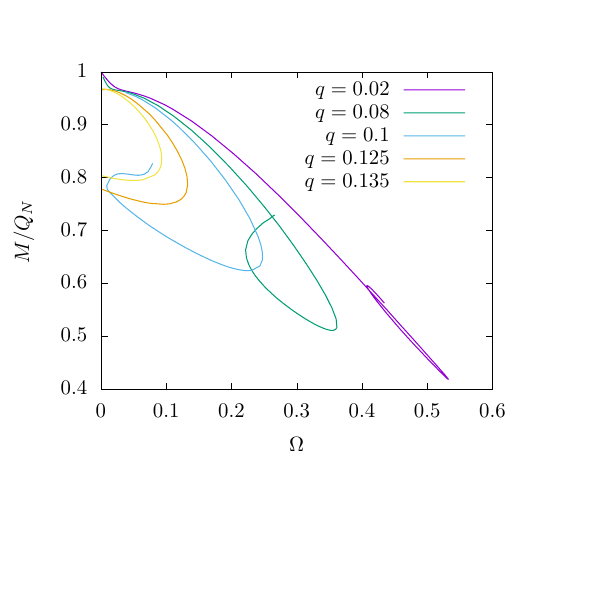}
\vspace{-2cm}
\caption{{\it Left}: The dependence of the mass $M$ on $\Omega$  for a massive scalar field with self-interaction, for $\alpha=0.012$ and for several values of $q$.
{\it Right}: The dependence of the ratio $M/Q_N$ on $\Omega$ for the same solutions. 
    \label{fig:mass_ratio_alp0_012}}
\end{figure}

We have also studied a larger value of $\alpha$ to understand the effect of stronger gravitational coupling on the results given above. Our results for
$\alpha=0.012$ are given in Fig.\ref{fig:mass_omega_sigma0_alp0_012}
and in Fig. \ref{fig:mass_ratio_alp0_012}. 
In this case, the value of $q$ which signals transition from sub- to super-criticality is $q_{\rm cr}\approx 0.0775$. 
Close to this value, i.e. at $q=0.08$, we find that three branches of solutions exist, however, that other than in the case $\alpha=0.0012$ no solutions with oscillations exist. The limiting phenomenon here is related to the fact that
$\sigma(0)\rightarrow 0$, i.e. the solution forms a space-time singularity at a finite central density $\psi(0)$.

Interestingly, while for $\alpha=0.0012$ we find that a change of pattern of solutions happens close
to the transition from sub- to super-criticality, this is no longer true for $\alpha=0.012$. For $q=0.1$ we still find three branches of solutions, while the phenomenon that solutions exist only on a finite interval of $\psi(0)$ with $\Omega=0$ at the two ends of the interval appears only for 
$q\gtrsim 0.125$. We did not manage to construct spatially oscillating solutions in this case.
Note that now all solutions are stable with respect to the decay into $Q_N$ individual bosons as $M/Q_N < 1$ for all solutions. We find that charged boson stars exist for  $q \leq 0.145$ when $\alpha=0.012$. \\

We have further studied the solutions for larger values of $\alpha$ and, as expected, 
we found that the larger $\alpha$ the larger $q$ can become, e.g. for $\alpha=1.0$ the solutions
can be constructed up to $q\approx 1.045$ with $\psi_{\rm min}\approx \psi_{\rm max}\approx 0.6$. In this case,
we find e.g. that at $q=1.0$ the branch of solutions exists for arbitrarily small values of $\psi(0)$, while for
$q=1.03$, we find that $\Omega=0$ at $\psi_{\rm min}(0)\approx 0.46$. Note that the critical value of $q$
is $q_{\rm cr}=1/\sqrt{2}$. 

\section{Black holes with charged scalar hair}
As discussed above, charged boson stars exist for $q\in [0:q_M]$. They correspond to uncharged boson stars in the limit $q=0$ and the maximal value of $q$, $q_M$, depends on $\alpha$. 
In the following, we will discuss the black hole counterparts to these solutions, i.e. black holes with charged scalar hair. Before discussing the details, let's recall that (i) black holes with uncharged scalar hair, i.e. for $q=0$ do not exist and (ii) black holes with charged scalar hair are known to exist in specific domains of the $\alpha$-$q$-plane. 

We need to impose boundary conditions at a regular horizon $r=r_h$. These are
\be
         N(r_h) = 0  \ \ , \ \ V(r_h) = 0   \ \ , \ \ N'\psi'\vert_{r=r_h} = \frac{1}{2}\frac{{\rm d} U_{(X)}}{{\rm d}\psi}
\ee
Note that $V'(r_h)\sim -E_r(r_h) \neq 0$, i.e. the radial electric field at the horizon does not vanish.
Essentially, the horizon presents an equipotential surface. The choice $V(r_h)=0$ stems from the
synchronization condition  $\omega - q V(r_h)=0$ in the gauge $\omega=0$ which we will use for black holes.

\subsection{Numerical results}
\begin{figure}[h]
\begin{center}
\includegraphics[width=8cm]{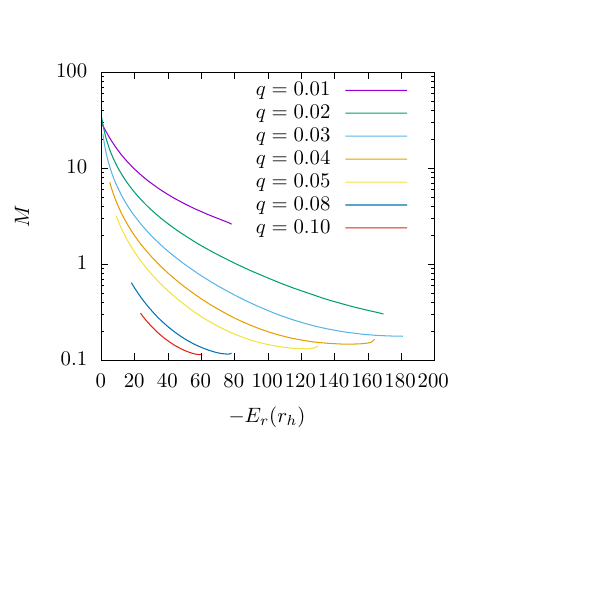}
\includegraphics[width=8cm]{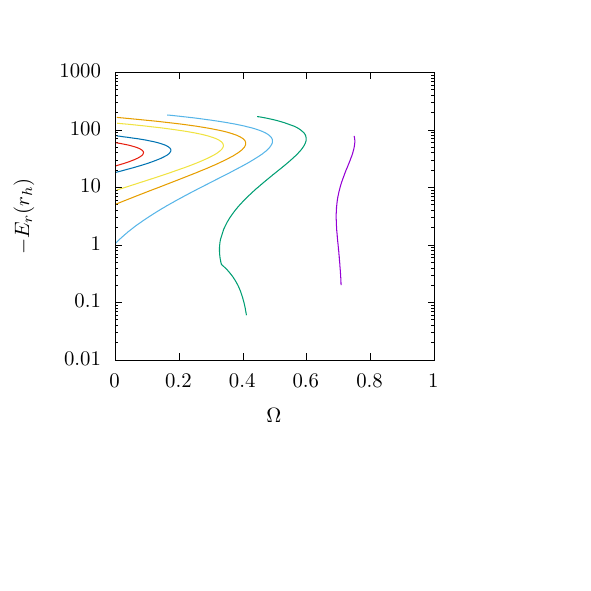}
\vspace{-2cm}\\

\includegraphics[width=8cm]{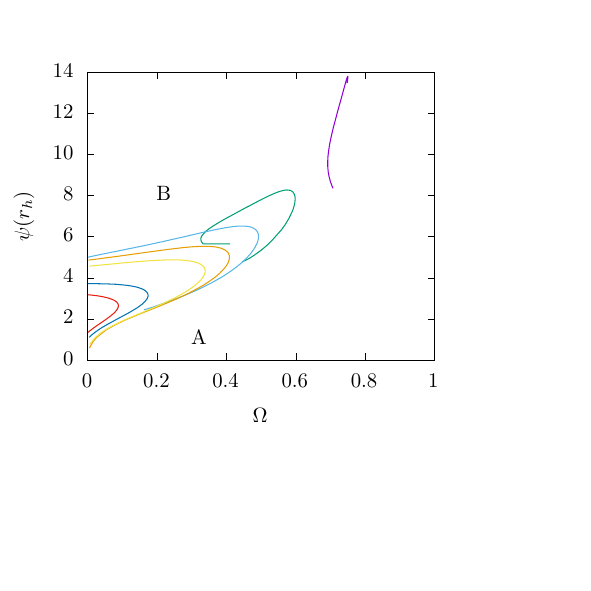}
\includegraphics[width=8cm]{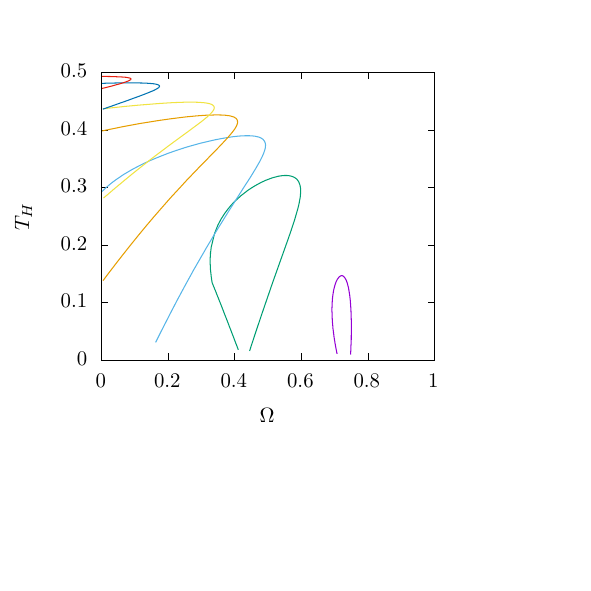}
\end{center}
\vspace{-2cm}
\caption{{\it Top left}: We show the mass $M$ as function of the value of the radial electric field at the horizon $-E_r(r_h)\sim V'(r_h)$ for black holes with charged scalar hair. {\it Top right}: We show the value of the radial electric field on the horizon $-E_r(r_h)\sim V'(r_h)$ as function of $\Omega$. {\it Bottom left}: We show the value of the scalar field on the horizon $\psi(r_h)$ as function of $\Omega$. {\it Bottom right}: We show the value of the Hawking temperature $T_H$  as function of $\Omega$. Here $\alpha= 0.0012$, $r_h = 0.15$ and $q=0.01$, $0.02$, $0.03$, $0.04$, $0.05$, $0.08$ and $0.1$, respectively - see caption in figure top left.
    \label{fig:BH_mass_Vp}}
\end{figure}

We have fixed the value of the event horizon to $r_h=0.15$ to study the pattern and compare it to the globally regular case. We believe that this choice of $r_h$ shows the generic features of the black hole solutions.
First of all, note that while for globally regular solutions $V'(0)=0$ is a regularity condition that states that there is no electric field at the center of the boson star, this is not longer true for black holes. The value of the radial electric field at the horizon can be varied within an interval that depends on
$\alpha$ and $q$, i.e. $V'(r_h)\sim -E_r(r_h) \in [V'_m:V'_M]$. This is shown in Fig. \ref{fig:BH_mass_Vp} (top left) for
$\alpha=0.0012$ and several values of $q$. For small values of $q$ we find that the electric field on the horizon can be made very small, albeit not zero as black holes with uncharged scalar hair do not exist. Moreover, $V'_m$ increases with increasing $q$. On the other hand, the maximal value of the electric field on the horizon $\sim V'_M$ increases only up to $q\approx 0.03$ and then starts to decrease again such that for $q\gtrsim 0.04$ black holes with charged scalar hair exist on smaller and smaller intervals of the strength of the electric field on the horizon.  
This demonstrates that black holes with charged scalar hair exist only when a subtle balance between the gravitational attraction, the scalar field attraction and repulsion as well as the electric repulsion between the individual 
bosons in the cloud around the black hole exists.

Fixing $\alpha$, we observe that black holes with charged scalar hair exist only on a finite domain of $q$, i.e.
for $q\in [q_m,q_M]$, where $q_m$ and $q_M$ depend on $\alpha$. E.g. we find that for $\alpha=0.0012$: $q_m\approx 0.005$, $q_M\approx 0.120$, while for $\alpha=0.035$ we get $q_m\approx 0.130$ and $q_M\approx 0.220$.
This indicates that the individual bosons making up the charged scalar cloud around the black holes need to carry a minimal, non-vanishing charge which increases with increased gravitational coupling (providing ``enough'' repulsion to counterbalance the attraction), while they cannot carry too much charge as in that case their repulsion would exceed the gravitational attraction and no localized solution should exist.  

In Fig.\ref{fig:BH_mass_Vp} (top right) we show the value of the radial electric field $-E_r(r_h)\sim V'(r_h)$ 
as function of $\Omega$ for $\alpha=0.012$, $r_h=0.15$ and several values of $q$. Note that $V(r_h)=0$ and hence the value of $v(\infty)$ corresponds to the potential difference between the horizon and infinity. As can be seen here, for large values of $q$, there exist two branches of solutions in $\Omega$, both ending at $\Omega=0$, i.e. where the potential difference between the horizon and infinity becomes so large that scalar particles of small $\mu\equiv 1$ can be produced. The two branches merge at $\Omega=\Omega_{\rm max}$ in this case. For large and increasing $q$ the value of $\Omega_{\rm max}$ decreases such that for sufficiently large $q$ charged scalar clouds don't exist anymore around the black hole. This can also be seen in Fig. \ref{fig:BH_mass_Vp} (bottom left) where we show the value of the scalar field at the horizon, $\psi(r_h)$, as function of $\Omega$. In this figure, we indicate the first branch (A) and the second branch (B). In fact, small scalar fields on the horizon allow for large radial electric fields on the horizon and vice versa. 

For intermediate values of $q$ (here $q=0.03$) we find that the A-branch of solutions stops at $\Omega=\tilde{\Omega} > 0$, while the B-branch still extends all the way back to $\Omega=0$. Decreasing $q$ further, we find that $\tilde{\Omega}$ increases, while branch B now also ends at $\Omega=\bar{\Omega} > 0$ and from there a third branch of solutions extends backwards in $\Omega$. On this third branch of solutions, the value of $-E_r(r_h)$ decreases
strongly (see top right). Moreover, the scalar field develops oscillations in the scalar field - for more details see \cite{Brihaye:2021mqk}. Finally, for $q=0.01$, we find that solutions only exist on a very small interval of $\Omega$. To understand these qualitative features better, we have studied the Hawking temperature $T_H$ of the solutions in dependence on $\Omega$, see Fig.\ref{fig:BH_mass_Vp} (bottom right). For all values of $q$ we find that the Hawking temperature increases from its value at the minimal possible value of $\Omega$ to a maximal value and then decreases again until the maximal possible value of $\Omega$. This decrease is connected to the formation of a plateau in $N(r)$ with value close to zero around the horizon $r_h$. Our numerical results further indicate that the limiting solution will be a solution that possesses an extremal horizon $r_h^{[ex)} > r_h$ with an extremal Reissner-Nordstr\"om solution forming outside this horizon and a non-trivial, scalarized solution for $r\in [r_h:r_h^{[ex)}]$. For $q=0.02$ and $q=0.01$ we observe that the Hawking temperature is small at both the minimal as well as at the maximal value of $\Omega$. Let us discuss the difference of these two small temperature solutions using the example $q=0.01$. In this case, the maximal value of $\Omega$ is $\Omega\approx 0.75$. The solution for this value of $\Omega$ has a large radial electric field on the horizon with $-E_r(r_h)\approx 78$ and small scalar field.
Moreover, $\sigma(r_h)\approx 0.48$
and $N(r_h)\sim (r-r_h)^n$ with $n > 1$. The solution also possesses a local minimum of the metric function $N(r)$ at $r\approx 10$ and we find that the scalar field function $\psi(r)\equiv 0$ for $r\gtrsim 10$. On the other hand the solution at the minimal value of $\Omega\approx 0.71$ has $-E_r(r_h)\approx 0.2$, i.e. much smaller electric field at the horizon, and a large scalar field. These qualitative features also appear for $q=0.02$, but additionally
we find that around the plateau formed for $N(r)$ oscillations of the scalar field function are possible.

\section{Conclusions}
When considering charged boson stars made off a self-interacting scalar field, we find that in contrast to the purely massive
case, super-critical boson stars can be stable to decay into their individual constituents. Moreover, these objects can exist for much larger values of $q$ up to $q_M$ than the argument that boson stars should only exist as long as 
the gravitational attraction between two charged scalar bosons dominates over the corresponding Coulomb repulsion would suggest. In fact, this is already true in the massive case, but the self-interaction enhances this effect, e.g. choosing $\alpha=1$ we find $q_M\approx 0.73995$ for the purely massive case, while  in the self-interacting case we have $q_M\approx 1.0450$, while the limit suggested by the balancing argument would be $q_{\rm cr}=1/\sqrt{2}$.

When the gravitational coupling is small, the change of the qualitative pattern of solutions happens roughly at $q=q_{\rm cr}$.
For $q \lesssim q_{\rm cr}$ solutions exist for arbitrarily small central density $\psi(0)$, while this is
no longer true for $q \gtrsim q_{\rm cr}$. In the latter case, the central density needs to be sufficiently large
for solutions to exist, i.e. super-critical boson stars need a sufficiently large scalar field density.
Just below the transition from sub- to super-criticality we observe that boson stars can possess spatial oscillations
outside their scalar core, a phenomenon that was first observed for charged boson stars with
a 6th order self-interaction scalar potential.

Increasing $\alpha$, i.e. the backreaction between the space-time and the matter content, we find that this is no longer necessarily true: super-critical boson stars can exist for arbitrarily small central density of the scalar field.



 \end{document}